\documentclass[preprint,showpacs,showkeys,preprintnumbers,amsmath,amssymb]{revtex4}
\usepackage{amssymb}

\begin{document}

\title{Symmetry group theorem to the Lin-Tsien equation and conservation laws relating
to the symmetry of the equation}

\author{ Xi-zhong LIU$^1$,  Qian-min QIAN$^1$}

\affiliation{$^1$ Department of Physics, Shaoxing College of Arts and Sciences, Shaoxing, 312000,
People¡¯s Republic of China} E-mail:
liuxizhong123@163.com
\begin{abstract}
We derive the symmetry group theorem to the Lin-Tsien equation by
using the modified CK's direct method, from which we obtain the
corresponding symmetry group. Conservation laws corresponding to the
Kac-Moody-Virasoro symmetry algebra of Lin-Tsien equation is
obtained up to second order group invariants.
\end{abstract}



\maketitle
\section{Introduction}
Conservation law which originates in mechanics and physics play an
important role in physics and mathematics. Nonlinear partial
differential equations (NPDEs) that admit conservation laws arise in
many disciplines of the applied sciences including physical
chemistry, fluid mechanics, particle and quantum physics, plasma
physics, elasticity, gas dynamics, electromagnetism,
magnetohydro-dynamics, nonlinear optics, and the bio-sciences.
Especially, in soliton theory, conservation laws have many
significant uses, particularly with regard to integrability and
linearization, analysis of solutions, and numerical solution
methods. Furthermore, completely integrable NPDEs
\cite{Ablowitz1,Ablowitz} admit infinitely many independent
conservation laws. Besides, finding the symmetry of NPDEs is also
very important. The mathematical foundations for the determination
of the full group for a system of differential equations can be
found in Ames \cite{Ames}, Bluman and Cole \cite{Bluman1}, and the
general theory is found in Ovsiannikov \cite{Ovsiannikov}. Among
them, the modified CK's direct method \cite{Ma,Lou} is an effective
method for finding symmetries and one advantage of this method is
that one can easily obtain the relationship between new exact
solutions and old ones of the given NPDEs.

Conservation law is closely connected with the the symmetry and this
connection is given by the famous Noether theorem. In the classical
Noether¡¯s theorem \cite{Noether}, if a given system of differential
equations has a variational principle, then a continuous symmetry
(point, contactor higher order) that leaves invariant the action
functional to within a divergence yields a conservation law
\cite{Bessel}-\cite{Bluman}. The Noether theorem has been the only
general device allowing one, in the class of Euler-Lagrange
equations, to reduce the search for conservation laws to a search
for symmetries.In the last few years, effective methods have been
devised for finding conservation laws for the very special class of
so-called Lax equations. In 2000, Kara \cite{KM} presented the
direct relationship between the conserved vector of a PDE and the
Lie-B\"{a}cklund symmetry generators of the PDE, from which it is
possible for us to obtain conservation laws from symmetries (see,
e.g. \cite{lou1}).

The Lin-Tsien equation \cite{Oswatitsch}
\begin{eqnarray}
2u_{tx}+u_xu_{xx}-u_{yy}=0, \label{kp}
\end{eqnarray}
where the variable $u\equiv u(x,y,t)$ is a velocity potential, is
known as a completely integrable model. It has been widely used to
study dynamic transonic flow in two space dimensions, plasma
physics, optics, condense matter physics, etc.

This paper is organized as follows. In section two, we derive the
symmetry group theorem to the Lin-Tsien equation by using the
modified CK's direct method, then the corresponding Lie point
symmetry groups and infinite dimensional Kac-Moody-Virasoro symmetry
algebra \cite{Kac} are obtained straightforwardly. As a comparison,
we also derive the Lie point symmetry groups by traditional Lie
approaches and the result shows that both methods produce the same
result. In section three, we first review some  basic notions about
Lie-B\"{a}cklund operators and then using it we finally obtain
conservation laws of the infinite dimensional Kac-Moody-Virasoro
symmetry algebra as Lin-Tsien equation possessed up to second order
group invariants. It is emphasized that equations with the same
symmetries may possesses the same types of conservation laws. The
last section is a short summary and discussion.
\section{Transformation Group by the Direct Method and
Kac-Moody Virasoro structure of the Lie point symmetry algebra}To
find complete point symmetry transformation group of (\ref{kp}), one
should find the general transformations in the following form
\begin{eqnarray}\label{fg0}
u=U(x,\ y,\ t,\ F(\xi, \eta, \tau)),
\end{eqnarray}
where $\xi,\ \eta$ and $\tau$ are functions of $ x,\ y,\ t$,
$F\equiv F(\xi, \eta, \tau)$ is also solutions of the Lin-Tsien
equation in the variables $\xi, \eta$ and $\tau$, i.e.,
\begin{eqnarray}\label{1F}
2F_{\tau,\xi}+F_\xi F_{\xi\xi}-F_{\eta\eta}=0,
\end{eqnarray}

Fortunately, we can prove that for the Lin-Tsien equation it is
enough to take
\begin{eqnarray}\label{fg}
u=\alpha+\beta F(\xi, \eta, \tau),
\end{eqnarray}
instead of (\ref{fg0}), where $\alpha$, $\beta$ and $\xi$, $\eta$,
$\tau$ are functions of $\{x, y, t\}$.

To prove the conclusion (\ref{fg}), one should submit the general
expression (\ref{fg0}) to Eq. (\ref{kp}). After eliminating
$F_{\eta\eta}$ and their higher derivatives via (\ref{1F}) and
vanishing all the coefficients of the different terms of the
derivatives of the functions $F$, one can get many complicated
determining equations for 4 functions $U\equiv U(x,\ y,\ t,\ F(\xi,
\eta, \tau)),\ \xi,\ \eta$ and $\tau$. Two of them read as
$$\xi_x\eta_x^2U_FF_\xi=0,\quad
\tau_x\xi_x^2U_FF_\tau=0,
$$
For the reason that $U_F$ should not be zero and there is no
nontrivial solution for $\xi_x = 0$, the only way to cause the
coefficients of $F_\xi$ and $F_\tau$ to vanish is

\begin{equation}\label{etax}
\eta_x=0,\quad \tau_x=0.
\end{equation}
Under the condition (\ref{etax}), the  determining equation
containing $F_\xi$ is
$$(U_F\xi_x^3-\eta_y^2)F_\xi=0.$$ \\
solving the above equation for $U$ then proves the conclusion that
assumption (\ref{fg}) instead of the general one (\ref{fg0}) is
sufficient to find the general symmetry group of the Lin-Tsien
equation.

Now the substitution of (\ref{fg}) with (\ref{etax}) into the
Lin-Tsien equation leads to
\begin{eqnarray}\label{w3}
&&(-2\beta_y\eta_y+2\beta_x\eta_t-\beta\eta_{yy})F_\eta+\xi_x\beta(\xi_{xx}
\beta+2\xi_x\beta_x)F_\xi^2+(\beta(\xi_x^3\beta+\tau_y\xi_y
\nonumber
\\&&-\tau_t\xi_x)F_{\xi\xi}+(\beta_x\beta\xi_{xx}+2\beta_x^2\xi_x
+\beta\xi_x\beta_{xx})F+2\beta\xi_{tx}+\alpha_x\beta\xi_{xx}\nonumber
\\&&+2\alpha_x\beta_x\xi_x
-2\beta_y\xi_y+\beta\xi_x\alpha_{xx}+2\beta_t\xi_x+2\beta_x\xi_t-\beta\xi_{yy}
)F_\xi-\beta\tau_y^2F_{\tau\tau}\nonumber
\\&&+(2\beta_x\tau_t-2\beta_y\tau_y
-\beta\tau_{yy})F_\tau+(\beta_x\beta\xi_x^2F+\beta(\alpha_x\xi_x^2
+2\xi_t\xi_x-\xi_y^2))F_{\xi\xi}\nonumber
\\&&+\beta(-\eta_y^2+\tau_t\xi_x-\tau_y\xi_y)F_{\eta\eta}
+2\beta(\eta_t\xi_x-\eta_y\xi_y)F_{\xi\eta}+\beta_x\beta_{xx}F^2
\nonumber\\&&+(2\beta_{tx}+\beta_x\alpha_{xx}+\alpha_x\beta_{xx}
-\beta_{yy})F+2\alpha_{tx}+\alpha_x\alpha_{xx} -\alpha_{yy}\nonumber
\\&&-2\beta\eta_y\tau_yF_{\tau\eta}=0.
\end{eqnarray}
Eq. (\ref{w3}) is true for arbitrary solutions $F$ only when all the
coefficients of the polynomials of the derivatives of $F$ are zero
which lead to a system of the determining equations for $\xi$,\
$\eta$, \ $\tau$, \ $\alpha$ and $\beta$
\begin{eqnarray}\label{c1}
-2\beta_y\eta_y+2\beta_x\eta_t-\beta\eta_{yy}=0,\quad
\xi_x\beta(\xi_{xx} \beta+2\xi_x\beta_x)=0,
\end{eqnarray}
\begin{eqnarray}
2\beta\xi_{tx}+\alpha_x\beta\xi_{xx}+2\alpha_x\beta_x\xi_x
-2\beta_y\xi_y+\beta\xi_x\alpha_{xx}+2\beta_t\xi_x+2\beta_x\xi_t-\beta\xi_{yy}=0,
\end{eqnarray}
\begin{eqnarray}
\beta_x\beta\xi_{xx}+2\beta_x^2\xi_x
+\beta\xi_x\beta_{xx}=0,\quad\beta\tau_y^2=0,\quad
2\beta_x\tau_t-2\beta_y\tau_y -\beta\tau_{yy}=0,
\end{eqnarray}
\begin{eqnarray}
(2\beta_{tx}+\beta_x\alpha_{xx}+\alpha_x\beta_{xx}
-\beta_{yy})=0,\quad\beta(-\eta_y^2+\tau_t\xi_x-\tau_y\xi_y)=0,
\end{eqnarray}
\begin{eqnarray}
\beta(\xi_x^3\beta+\tau_y\xi_y -\tau_t\xi_x)=0,\quad
\beta_x\beta\xi_x^2=0,\quad 2\beta(\eta_t\xi_x-\eta_y\xi_y)=0,
\end{eqnarray}
\begin{eqnarray}
\beta_x\beta_{xx}=0,\quad\beta(\alpha_x\xi_x^2
+2\xi_t\xi_x-\xi_y^2)=0,\quad2\alpha_{tx}+\alpha_x\alpha_{xx}
-\alpha_{yy}=0,
\end{eqnarray}
\begin{eqnarray}\label{c2}
\beta\eta_y\tau_y=0.
\end{eqnarray}
It is straightforward to obtain the general solutions of the
determining equations (\ref{c1})--(\ref{c2}). The results are

\begin{eqnarray}
&&\xi =
\tau_t^{\frac{1}{3}}x+\frac{1}{3}\frac{y^2\tau_{tt}}{\tau_t^{\frac{2}{3}}}
+\frac{y\eta_{0t}}{\tau_t^{\frac{1}{3}}}+\xi_0, \eta =
\tau_t^{\frac{2}{3}}y+\eta_0, \beta =
\tau_t^{\frac{1}{3}},\label{taubeta}
\\&&\alpha=-\frac{(-36\tau_{tt}\tau_{ttt}\tau_t+28\tau_{tt}^3
+9\tau_{tttt}\tau_t^2)y^4}{81\tau_t^3}
\nonumber\\&&-\frac{2(-2\eta_{0tt}\tau_{tt}\tau_t
+\eta_{0ttt}\tau_t^2-\tau_{ttt}\eta_{0t}\tau_t
+2\tau_{tt}^2\eta_{0t})y^3}{3\tau_t^{\frac{8}{3}}}\nonumber\\&&+(-\frac{2(-4\tau_{tt}
^2+3\tau_{ttt}\tau_t)x}{9\tau_t^2}-\frac{-4\tau_{tt}\xi_{0t} \tau_t
-6\eta_{0t}\eta_{0tt}\tau_t+6\xi_{0tt}\tau_t^2+5\tau_{tt}\eta_{0t}^2}{3\tau_t^{\frac{7}{3}}}
)y^2
\nonumber\\&&+(-\frac{2(-\eta_{0t}\tau_{tt}+\eta_{0tt}\tau_t)x}{\tau_t^{\frac{5}{3}}}+\alpha_2)y
-\frac{\tau_{tt}x^2}{3\tau_t}-\frac{(2\xi_{0t}\tau_t-\eta_{0t}^2)x}{\tau_t^{\frac{4}{3}}}
+\alpha_1
\end{eqnarray}
where $\xi_0 \equiv \xi_0(t )$, $\eta_0 \equiv \eta_0(t)$, $\tau
\equiv \tau(t)$, $\alpha_1\equiv\alpha_1(t)$ and
$\alpha_2\equiv\alpha_2(t)$ are arbitrary functions of time t.

 In summary, the following theorem holds:\\
\textbf{Theorem 1:} If $F=F(x,y,t)$ is a solution of the Lin-Tsien
equation (\ref{kp}), then so is
\begin{eqnarray}\label{th1}
u&=&-\frac{(-36\tau_{tt}\tau_{ttt}\tau_t+28\tau_{tt}^3
+9\tau_{tttt}\tau_t^2)y^4}{81\tau_t^3}
\nonumber\\&&-\frac{2(-2\eta_{0tt}\tau_{tt}\tau_t
+\eta_{0ttt}\tau_t^2-\tau_{ttt}\eta_{0t}\tau_t
+2\tau_{tt}^2\eta_{0t})y^3}{3\tau_t^{\frac{8}{3}}}\nonumber\\&&+(-\frac{2(-4\tau_{tt}
^2+3\tau_{ttt}\tau_t)x}{9\tau_t^2}-\frac{-4\tau_{tt}\xi_{0t} \tau_t
-6\eta_{0t}\eta_{0tt}\tau_t+6\xi_{0tt}\tau_t^2+5\tau_{tt}\eta_{0t}^2}{3\tau_t^{\frac{7}{3}}}
)y^2
\nonumber\\&&+(-\frac{2(-\eta_{0t}\tau_{tt}+\eta_{0tt}\tau_t)x}{\tau_t^{\frac{5}{3}}}+\alpha_2)y
-\frac{\tau_{tt}x^2}{3\tau_t}-\frac{(2\xi_{0t}\tau_t-\eta_{0t}^2)x}{\tau_t^{\frac{4}{3}}}
\nonumber\\&&+\alpha_1+\tau_t^{\frac{1}{3}}F(\xi,\eta,\tau),
\end{eqnarray}
with (\ref{taubeta}), where $\xi_0$, $\eta_0$, $\tau$, $\alpha_1$
and $\alpha_2$ are arbitrary functions of $t$.

Applying the theorem to some simple exact solutions without
arbitrary functions, one may obtain some types of novel generalized
solutions with some arbitrary functions. In the following, we just
present one
 special solution example.\\
\bf Example 1. \rm It is quite trivial that the Lin-Tsien equation
(\ref{kp}) possesses a special simple solution
\begin{eqnarray}
&&F=1. \label{F0}
\end{eqnarray}

Using the transformation theorem to the above special solution we
have the following new special solution of the Lin-Tsien equation
\begin{eqnarray}\label{F1}
u&=&-\frac{(-36\tau_{tt}\tau_{ttt}\tau_t+28\tau_{tt}^3
+9\tau_{tttt}\tau_t^2)y^4}{81\tau_t^3}
\nonumber\\&&-\frac{2(-2\eta_{0tt}\tau_{tt}\tau_t
+\eta_{0ttt}\tau_t^2-\tau_{ttt}\eta_{0t}\tau_t
+2\tau_{tt}^2\eta_{0t})y^3}{3\tau_t^{\frac{8}{3}}}\nonumber\\&&+(-\frac{2(-4\tau_{tt}
^2+3\tau_{ttt}\tau_t)x}{9\tau_t^2}-\frac{-4\tau_{tt}\xi_{0t} \tau_t
-6\eta_{0t}\eta_{0tt}\tau_t+6\xi_{0tt}\tau_t^2+5\tau_{tt}\eta_{0t}^2}{3\tau_t^{\frac{7}{3}}}
)y^2
\nonumber\\&&+(-\frac{2(-\eta_{0t}\tau_{tt}+\eta_{0tt}\tau_t)x}{\tau_t^{\frac{5}{3}}}+\alpha_2)y
-\frac{\tau_{tt}x^2}{3\tau_t}-\frac{(2\xi_{0t}\tau_t-\eta_{0t}^2)x}{\tau_t^{\frac{4}{3}}}
\nonumber\\&&+\alpha_1+\tau_t^{\frac{1}{3}}.
\end{eqnarray}

In the traditional Lie group theory, one always tries to find the
Lie point symmetries first and then use the Lie's first fundamental
theorem to obtain the symmetry transformation group. Conversely, we
are fortunate to obtain the symmetry transformation group in the
first place by a simple direct method. Once the transformation group
is known, the Lie point symmetries and the related Lie symmetry
algebra can be obtained straightforward by a more simple limiting
procedure.

For the Lin-Tsien (\ref{kp}), the corresponding Lie point symmetries
can be derived from the symmetry group transformation theorem by
setting
\begin{eqnarray} &&\eta_0(t)=\epsilon h(t), \quad
\tau(t)=t+\epsilon f(t), \quad \xi_0(t) = \epsilon g(t), \quad
\alpha_1(t) = \epsilon m(t), \nonumber\\&& \alpha_2(t)=\epsilon
n(t),
\end{eqnarray}
with $\epsilon$ being an infinitesimal parameter, then (\ref{th1})
can be written as
\begin{eqnarray}\label{sym1}
u&=&F+ \epsilon \sigma(F) + O(\epsilon^2), \nonumber \\
\sigma(F)&=&
(g(t)+\frac{1}{3}f_t(t)x+\frac{1}{3}f_{tt}(t)y^2+yh_t(t))F_x
+(h(t)+\frac{2}{3}f_t(t)y)F_y\nonumber
\\&&+fF_t+\frac{1}{3}f_t(t)F-\frac{2}{3}f_{ttt}xy^2
-\frac{1}{3}f_{tt}x^2-2g_t(t)x+m(t)\nonumber
\\&&+n(t)y-\frac{1}{9}f_{tttt}y^4
-\frac{2}{3}h_{ttt}(t)y^3-2g_{tt}y^2-2h_{tt}xy,
\end{eqnarray}

The equivalent vector expression of the above symmetry reads
\begin{eqnarray}
V&=&\{(\frac{1}{3}f_t(t)x+\frac{1}{3}f_{tt}(t)y^2)\frac{\partial}{\partial
x}+\frac{2}{3}f_t(t)y\frac{\partial}{\partial
y}+f(t)\frac{\partial}{\partial
t}\nonumber\\&&-(\frac{1}{3}f_t(t)F-\frac{2}{3}f_{ttt}xy^2
-\frac{1}{3}f_{tt}x^2-\frac{1}{9}f_{tttt}y^4)\frac{\partial}{\partial
F}\}\nonumber\\&&+\{g(t)\frac{\partial}{\partial
x}+(2g_t(t)x+2g_{tt}y^2)\frac{\partial}{\partial
F}\}+\{yh_t(t)\frac{\partial}{\partial
x}+h(t)\frac{\partial}{\partial
y}\nonumber\\&&+(\frac{2}{3}h_{ttt}(t)y^3+2h_{tt}xy)\frac{\partial}{\partial
F}\}-\{m(t)\frac{\partial}{\partial
F}\}-\{n(t)y\frac{\partial}{\partial F}\}\nonumber\\&&\equiv
V_1(f(t)) + V_2(g(t)) + V_3(h(t))+V_4(m(t))+V_5(n(t)).
\end{eqnarray}
Since the functions $f$, $g$, $h$, $m$ and $n$ are arbitrary, the
corresponding Lie algebra is an infinite dimensional Lie algebra.

It is easy to verify that the symmetries $V_i,\ i=1,\ 2,\ 3,\ 4,\ 5$
constitute an infinite dimensional Kac-Moody-Virasoro \cite{Kac}
type symmetry algebra {\bf \em S} with the following nonzero
commutation relations
\begin{eqnarray}
&&[V_1(f),\ V_4(m)]=V_4(\frac{1}{3}mf_t+fm_t),\label{alg11}\\
&&[V_2(g),\
V_3(h)]=V_5(2(2hg_{tt}-gh_{tt}+h_tg_t)),\label{alg12}\\
&&[V_1(f),\ V_3(h)]=V_3(fh_t-\frac{2}{3}hf_t),\label{alg13}\\
&&[V_1(f),\ V_2(g)]=V_2(g_tf-\frac{1}{3}gf_t)),\label{alg11}\\
&&[V_2(g_1),\
V_2(g_2)]=V_4(2g_{1t}g_2-2g_1g_{2t}),\label{alg12}\\
&&[V_3(h),\ V_5(n)]=V_4(hn),\label{alg13}\\
&&[V_1(f),\ V_5(n)]=V_5(f_tn+fn_t),\label{alg11}\\
&&[V_1(f_1),\
V_1(f_2)]=V_1(f_1f_{2t}-f_2f_{1t}),\label{alg12}\\
&&[V_3(h_1),\ V_3(h_2)]=V_2(h_1h_{2t}-h_2h_{1t}),\label{alg13}
\end{eqnarray}

It should be emphasized that the algebra is infinite dimensional
because the generators $V_1,\ V_2,\ V_3,\ V_4$ and $\ V_5$ all
contain \em arbitrary \rm functions. The algebra is closed because
all the commutators can be expressed by the generators belong to the
generator set usually with different functions and the generators
contained \em different \rm functions belong to the set. Especially,
it is clear that the symmetry $V_1(f)$ constitute an centerless
Virasoro symmetry algebra.

As a comparison, we now derive the Lie point symmetry of the
Lin-Tsien equation by the standard Lie approach briefly.

To study the symmetry of the equation (\ref{kp}), we search for the
Lie point symmetry transformations in the vector form
\begin{eqnarray}\nonumber
V=X \frac{\partial}{\partial x}+Y \frac{\partial}{\partial y}+T
\frac{\partial}{\partial t}+U \frac{\partial}{\partial u},
\end{eqnarray}
where $X$, $Y$, $T$ and $U$ are functions with respect to $x$, $y$,
$t$, $u$, which means that (\ref{kp}) is invariant under the point
transformation
\begin{eqnarray}
\{x, y, t, u\} \rightarrow \{x+\epsilon X, y+\epsilon Y, t+\epsilon
T, u+\epsilon U\} \nonumber
\end{eqnarray}
with infinitesimal parameter $\epsilon$.

In other words, the symmetry of the equation (\ref{kp}) can be
written as the function form
\begin{eqnarray}
\sigma=X u_x+Y u_y+T u_t-U, \label{sigma}
\end{eqnarray}
where the symmetry $\sigma$ is a solution of the linearized equation
for (\ref{kp})
\begin{eqnarray}
2\sigma_{tx}+ \sigma_{x} u_{xx}+u_{x} \sigma_{xx}-\sigma_{yy}=0,
\label{sigma1}
\end{eqnarray}
which is obtained by substituting $u=u+\epsilon \sigma$ into
(\ref{kp}) and dropping the nonlinear terms in $\sigma$.

It is easy to solve out $X(x,y,t,u)$, $Y(x,y,t,u)$, $T(x,y,t,u)$ and
$U(x,y,t,u)$ by substituting (\ref{sigma}) into (\ref{sigma1}), and
eliminating $u_{yy}$ and its higher order derivatives by means of
the Lin-Tsien equation. After taking the constants as zero, we get
the results
\begin{eqnarray}
X(x,y,t,u)=\frac{1}{3}T_{t}x+\frac{1}{3}T_{t t} y^2
  +X_{t} y+Y,
\end{eqnarray}
\begin{eqnarray}
Y(x,y,t,u)=\frac{2}{3}T_{t}y+X,
\end{eqnarray}
\begin{eqnarray}
T(x,y,t,u)=T(t),
\end{eqnarray}
\begin{eqnarray}
U(x,y,t,u)&=&\frac{1}{3} x^2 T_{tt}-\frac{1}{3} u T_t+\frac{2}{3} x
y^2 T_{ttt}+2 x X_{tt} y+2 x Y_t\nonumber \\
&&+\frac{1}{9}y^4
T_{tttt}+\frac{2}{3}X_{ttt}y^3+2Y_{tt}y^2+Z_1y+Z_2,
\end{eqnarray}
where $X$, $Y$, $T$, $Z_1$ and $Z_2$ are arbitrary functions of $t$.

The vector form of the Lie point symmetries reads
\begin{eqnarray} V&=&
\left(\frac{1}{3}T_{t}+\frac{1}{3}T_{t t} y^2
  +X_{t} y+Y
\right) \frac{\partial}{\partial x}+\left( \frac{2}{3}T_{t}+X
\right)\frac{\partial}{\partial y}+T \frac{\partial}{\partial t}
 \nonumber \\ &&+\big(\frac{1}{3} u T_t-\frac{1}{3} x^2 T_{tt}-\frac{2}{3} x y^2
T_{ttt}-2 x X_{tt} y-2 x Y_t-\frac{1}{9}y^4
T_{tttt}\nonumber\\&&-\frac{2}{3}X_{ttt}y^3-2Y_{tt}y^2-Z_1y-Z_2
\big)\frac{\partial}{\partial u},\label{V}
\end{eqnarray}
which is exactly the same as that obtained by the modified CK's
approach.

\section{Conservation Laws related to the the symmetry (\ref{V})}

In order to obtain conservation laws related to the symmetry
(\ref{V}), we need some  basic notions about Lie-B\"{a}cklund
operators first.

A Lie-B\"{a}cklund operator is given by
\begin{eqnarray}
X_0=\xi^i \frac{\partial}{\partial x^i}+\eta \frac{\partial
}{\partial u}+\zeta_i \frac{\partial}{\partial u_i}+\zeta_{i_1 i_2}
\frac{\partial}{\partial u_{i_1 i_2}}+\cdots, \label{LB}
\end{eqnarray}
where $\xi^i$, $\eta$ and the additional coefficients are
\begin{eqnarray}
&& \zeta_i=D_i (W)+\xi^j u_{i j}, \nonumber \\
&& \zeta_{i_1 i_2}=D_{i_1 i_2} (W)+\xi^j u_{j i_1 i_2},
\end{eqnarray}
and $W$ is the Lie characteristic function defined by
\begin{eqnarray}
W=\eta- \xi^j u_j
\end{eqnarray}
with $D_i$ being the operator of total differentiation
\begin{eqnarray}
D_i=\frac{\partial}{\partial x^i}+u_i \frac{\partial}{\partial
u}+u_{ij}\frac{\partial}{\partial u_j}+\cdots, \qquad  i=1, \cdots,
n,
\end{eqnarray}
as
\begin{eqnarray}
u_i= D_i(u), \qquad u_{ij}= D_j D_i (u). \label{D}
\end{eqnarray}
These definitions and results relating to Lie-B\"{a}cklund operator
can be found in \cite{LB} and the repeated indices mean the
summations known as the Einstein summation rule.

Using equations (\ref{LB})-(\ref{D}), we can calculate the
$2$nd-order Lie-B\"{a}cklund operator of the vector field $V$
defined by equation (\ref{V}):
\begin{eqnarray}
\xi^x=\frac{1}{3}xT_{t}+\frac{1}{3}T_{t t} y^2+X_{t} y+Y,
\end{eqnarray}
\begin{eqnarray}
\xi^y=\frac{2}{3}yT_{t}+X,
\end{eqnarray}
\begin{eqnarray}
\xi^t= T,
\end{eqnarray}
\begin{eqnarray}
\eta&=&-\frac{1}{3} u T_t+\frac{1}{3} x^2 T_{tt}+\frac{2}{3} x y^2
T_{ttt}+2 x X_{tt} y+2 x Y_t+\frac{1}{9}y^4
T_{tttt}\nonumber \\
&&+\frac{2}{3}X_{ttt}y^3+2Y_{tt}y^2+Z_1y+Z_2,
\end{eqnarray}
\begin{eqnarray}
\zeta_x=-\frac{2}{3}T_t
u_x+\frac{2}{3}xT_{tt}+\frac{2}{3}y^2T_{ttt}+2X_{tt}y+2Y_t,
\end{eqnarray}
\begin{eqnarray}
\zeta_y&=&-T_t
u_y+\frac{4}{3}T_{ttt}yx+2xX_{tt}+\frac{4}{9}y^3T_{tttt}+2X_{ttt}y^2+4Y_{tt}y
\nonumber\\&&+Z_1-(\frac{2}{3}T_{tt} y+X_t)u_x,
\end{eqnarray}
\begin{eqnarray}
\zeta_t &=&-\frac{4}{3}T_t u_t-\frac{1}{3}u T_{tt}+\frac{1}{3}x^2
T_{ttt}+\frac{2}{3}x y^2 T_{tttt}+2xy
X_{ttt}+2xY_{tt}\nonumber \\
&&+\frac{1}{9}y^4T_{ttttt}+\frac{2}{3}X_{tttt}y^3+2Y_{ttt}y^2
+Z_{1,t}y+Z_{2,t}\nonumber\\&&-\big(\frac{1}{3}xT_{tt}
+\frac{1}{3}y^2 T_{ttt}+y X_{tt}+Y_t \big )u_x-\big(\frac{2}{3} y
T_{tt}+X_t \big)u_y,
\end{eqnarray}
\begin{eqnarray}
\zeta_{xx}=-u_{xx}T_t+\frac{2}{3}T_{tt},
\end{eqnarray}
\begin{eqnarray}
\zeta_{xy}=-\frac{4}{3}T_t
u_{xy}+\frac{4}{3}yT_{ttt}+2X_{tt}-\big(\frac{2}{3}y T_{tt}+X_t
\big)u_{xx},
\end{eqnarray}
\begin{eqnarray}
\zeta_{xt}&=&-\frac{5}{3}T_t
u_{xt}-\frac{2}{3}T_{tt}u_x+\frac{2}{3}xT_{ttt}+\frac{2}{3}y^2T_{tttt}
+2X_{ttt}y+2Y_{tt}\nonumber \\
&&-(\frac{1}{3}xT_{tt}+\frac{1}{3}y^2 T_{ttt}+y
X_{tt}+Y_t)u_{xx}-(\frac{2}{3}T_{tt} y+X_t)u_{xy},
\end{eqnarray}
\begin{eqnarray}
\zeta_{yy}&=&-\frac{5}{3}T_t
u_{yy}+\frac{4}{3}xT_{ttt}+\frac{4}{3}y^2T_{tttt}+4X_{ttt}y+4Y_{tt}-\frac{2}{3}
u_x T_{tt}\nonumber\\&&-2(\frac{2}{3}yT_{tt}+X_t)u_{xy},
\end{eqnarray}
\begin{eqnarray}
\zeta_{yt}&=&-2T_t
u_{yt}-T_{tt}u_y+\frac{4}{3}xyT_{tttt}+2xX_{ttt}+\frac{4}{9}y^3T_{ttttt}
+2X_{tttt}y^2\nonumber \\&&+4yY_{ttt}+Z_{1,t}-(\frac{2}{3}
yT_{ttt}+X_{tt})u_x -(\frac{2}{3}yT_{tt}+X_t)u_{xt}\nonumber
\\&&-(\frac{1}{3}xT_{tt}+\frac{1}{3}y^2T_{ttt}+yX_{tt}+Y_t)
u_{xy}-(\frac{2}{3}yT_{tt}+X_t)u_{yy},
\end{eqnarray}
\begin{eqnarray}
\zeta_{tt}&=&-\frac{7}{3}T_tu_{tt}-\frac{5}{3}T_{tt}
u_t-\frac{1}{3}T_{ttt} u+\frac{1}{3} x^2
T_{tttt}+\frac{2}{3}xy^2T_{ttttt}+2xy X_{tttt}\nonumber
\\&&+2xY_{ttt}+\frac{1}{9}y^4T_{tttttt}+\frac{2}{3}X_{ttttt}y^3+2Y_{tttt}y^2
+Z_{1,tt}y+Z_{2,tt}\nonumber \\&&-(\frac{1}{3} xT_{ttt}+\frac{1}{3}
y^2 T_{tttt}+y X_{ttt}+Y_{tt})u_x-2(\frac{1}{3}x
T_{tt}+\frac{1}{3}y^2 T_{ttt}\nonumber
\\&&+y
X_{tt}+Y_t)u_{xt}-(\frac{2}{3}yT_{ttt}+X_{tt})u_y-2(\frac{2}{3}y
T_{tt}+X_t)u_{yt}.
\end{eqnarray}
Correspondingly, the second order Lie-B\"{a}cklund operator is given
by
\begin{eqnarray}
X_0&=&\xi^x \frac{\partial}{ \partial x}+\xi^y \frac{\partial}{
\partial y}+\xi^t \frac{\partial}{ \partial t}+\eta \frac{\partial}{\partial u}
+\zeta_x \frac{\partial}{ \partial u_x} +\zeta_y \frac{\partial}{
\partial u_y}+\zeta_t \frac{\partial}{\partial u_t}+\zeta_u \frac{\partial}{\partial u_{xx}}
\nonumber \\
&&+\zeta_{xy} \frac{\partial}{\partial u_{xy}} +\zeta_{xt}
\frac{\partial}{\partial u_{xt}}+\zeta_{yy} \frac{\partial}{\partial
u_{yy}}+\zeta_{yt} \frac{\partial}{\partial u_{yt}}+\zeta_{tt}
\frac{\partial}{\partial u_{tt}}. \label{x0}
\end{eqnarray}

\textbf{Theorem}(\cite{cl}, \cite{KM}): Suppose that $X_0$ is a
Lie-B\"{a}cklund symmetry of (\ref{kp}) such that the conservation
vector $T'=(T_1, \ T_2, \ T_3)$ is invariant under $X_0$. Then
\begin{eqnarray}
X_0(T_i)+\sum_{j=1}^3 T_i D_j{\xi_j}-\sum_{j=1}^3 T_j D_j (\xi_i)=0,
\quad (i=1,\ 2,\ 3), \label{cl}
\end{eqnarray}
where $D_1=D_x$, $D_2=D_y$, $D_3=D_t$ and $\xi_i$ are determined by
(\ref{x0}).

A Lie-B\"{a}cklund symmetry $X_0$ is said to be associated with a
conserved vetor $T'$ of (\ref{kp}) if $X_0$ and $T'$ satisfy
relations (\ref{cl}).

Now we construct the corresponding conservation laws relating to
(\ref{V}) in the form
\begin{eqnarray}
D_x J_1+D_y J_2+D_t \rho=0, \label{cl0}
\end{eqnarray}

where $T_1=J_1$, $T_2=J_2$, $T_3=\rho$ with $J_1$, $J_2$ and $\rho$
being functions of \{$x$, $y$, $t$, $u$, $u_x$, $u_y$, $\cdots$,
$u_{tt}$\}.

In terms of $T'=(J_1, \ J_2, \rho)$, Eq. (\ref{cl}) is equivalent to
the following three equations:
\begin{eqnarray}
&&(\frac{1}{3}xT_t+\frac{1}{3}y^2T_{tt}+yX_t+Y)\frac{\partial
J_1}{\partial x} +(\frac{2}{3}y T_t+X )\frac{\partial J_1}{\partial
y} +T\frac{\partial J_1}{\partial t}
+(-\frac{1}{3}uT_t\nonumber \\
&&+\frac{1}{3}x^2T_{tt}+\frac{2}{3}xy^2 T_{ttt}+2xy
X_{tt}+2xY_t+\frac{1}{9}y^4T_{tttt}+\frac{2}{3}X_{ttt}y^3+2Y_{tt}y^2\nonumber \\
&&+Z_1y+Z_2)\frac{\partial J_1}{\partial u}+(-\frac{2}{3}T_t
u_x+\frac{2}{3}xT_{tt}+\frac{2}{3}y^2T_{ttt}+2X_{tt}y+2Y_t)\frac{\partial
J_1}{\partial u_x}\nonumber \\ &&+\left[-T_t u_y+\frac{4}{3}x y
T_{ttt}+2x X_{tt}+\frac{4}{9}y^3T_{tttt}+2X_{ttt}y^2+4Y_{tt}y+Z_1
\right. \nonumber \\
&& \left.-(\frac{2}{3} y T_{tt}+X_t)u_x\right]\frac{\partial
J_1}{\partial u_y} +\left[-\frac{4}{3} T_t u_t-\frac{1}{3}
uT_{tt}+\frac{1}{3} x^2 T_{ttt}+\frac{2}{3}xy^2T_{tttt}\right.\nonumber \\
&&\left.+2x yX_{ttt}+2xY_{tt}+\frac{1}{9}y^4T_{ttttt}
+\frac{2}{3}X_{tttt}y^3+2Y_{ttt}y^2 +Z_{1,t}y+Z_{2,t}\right.\nonumber\\
&&\left.-(\frac{1}{3} x T_{tt}+\frac{1}{3}y^2 T_{ttt}+y
X_{tt}+Y_{t}) u_x -(\frac{2}{3} y
T_{tt}+X_t)u_y)\right]\frac{\partial J_1}{\partial u_t} +(-T_t
u_{xx}\nonumber\\
&&+\frac{2}{3}T_{tt})\frac{\partial J_1}{\partial
u_{xx}}+\left[-\frac{4}{3} T_t u_{xy}+\frac{4}{3}y T_{ttt}
+2X_{tt}-(\frac{2}{3}yT_{tt}+X_t)u_{xx} \right]\frac{\partial
J_1}{\partial u_{xy}}\nonumber\\&&+\left[-\frac{5}{3}T_t
u_{xt}-\frac{2}{3}T_{tt} u_x+\frac{2}{3}
xT_{ttt}+\frac{2}{3}y^2T_{tttt}+2X_{ttt}y+2Y_{tt}-(\frac{1}{3} x
T_{tt}\right.\nonumber\\
&&\left.+\frac{1}{3} y^2 T_{ttt}+y X_{tt} +Y_t)u_{xx}-(\frac{2}{3}y
T_{tt}+X_t)u_{xy}\right]\frac{\partial J_1}{\partial
u_{xt}}+\left[-\frac{5}{3}T_tu_{yy}+\frac{4}{3}xT_{ttt}\right. \nonumber\\
&&\left.+\frac{4}{3}y^2T_{tttt}
+4X_{ttt}y+4Y_{tt}-\frac{2}{3}u_xT_{tt}-2(\frac{2}{3}yT_{tt}+X_t)u_{xy}\right]\frac{\partial
J_1}{\partial
u_{yy}}\nonumber\\&&+\left[-2u_{yt}T_t-u_yT_{tt}+\frac{4}{3}xyT_{tttt}+2xX_{ttt}+\frac{4}{9}y^3T_{ttttt}
+2X_{tttt}y^2+4Y_{ttt}y\right. \nonumber\\
&&\left.+Z_{1,t}-(\frac{2}{3}yT_{ttt}+X_{tt})u_x-(\frac{2}{3}yT_{tt}
+X_t)u_{tx}-(\frac{1}{3}xT_{tt}+\frac{1}{3}y^2T_{ttt}+X_{tt}y\right. \nonumber\\
&&\left.+Y_t)u_{xy}-(\frac{2}{3}yT_{tt}
+X_t)u_{yy}\right]\frac{\partial J_1}{\partial
u_{yt}}+\left[-\frac{7}{3}u_{tt}T_t-\frac{5}{3}u_tT_{tt}-\frac{1}{3}uT_{ttt}
\right. \nonumber\\
&&\left.+\frac{1}{3}x^2T_{tttt}+\frac{2}{3}xy^2T_{ttttt}+2xX_{tttt}y+2xY_{ttt}
+\frac{1}{9}y^4T_{tttttt}+\frac{2}{3}X_{ttttt}y^3+2Y_{tttt}y^2\right.\nonumber\\
&&\left.+Z_{1,tt}y
+Z_{2,tt}-(\frac{1}{3}xT_{ttt}+\frac{1}{3}y^2T_{tttt}+X_{ttt}y+Y_{tt})u_x-2(\frac{1}{3}
xT_{tt}+\frac{1}{3}y^2T_{ttt}\right.\nonumber\\
&&\left.+X_{tt}y+Y_t)u_{tx}-(\frac{2}{3}yT_{ttt}
+X_{tt})u_y-2(\frac{2}{3}yT_{tt}+X_t)u_{yt}\right]\frac{\partial
J_1}{\partial u_{tt}}+\frac{5}{3} T_t J_1\nonumber\\
&& -(X_t+\frac{2}{3} y T_{tt})J_2-(\frac{1}{3} x T_{tt}+\frac{1}{3}
y^2 T_{ttt}+y X_{tt}+Y_t)\rho=0, \label{cl1}
\end{eqnarray}
\begin{eqnarray}
&&(\frac{1}{3}xT_t+\frac{1}{3}y^2T_{tt}+yX_t+Y)\frac{\partial
J_2}{\partial x} +(\frac{2}{3}y T_t+X )\frac{\partial J_2}{\partial
y} +T\frac{\partial J_2}{\partial t}
+(-\frac{1}{3}uT_t\nonumber \\
&&+\frac{1}{3}x^2T_{tt}+\frac{2}{3}xy^2 T_{ttt}+2xy
X_{tt}+2xY_t+\frac{1}{9}y^4T_{tttt}+\frac{2}{3}X_{ttt}y^3+2Y_{tt}y^2\nonumber \\
&&+Z_1y+Z_2)\frac{\partial J_2}{\partial u}+(-\frac{2}{3}T_t
u_x+\frac{2}{3}xT_{tt}+\frac{2}{3}y^2T_{ttt}+2X_{tt}y+2Y_t)\frac{\partial
J_2}{\partial u_x}\nonumber\\&&+\left[-T_t u_y+\frac{4}{3}x y
T_{ttt}+2x X_{tt}+\frac{4}{9}y^3T_{tttt}+2X_{ttt}y^2+4Y_{tt}y+Z_1
 \right. \nonumber \\ && \left.-(\frac{2}{3} y T_{tt}+X_t)u_x\right]\frac{\partial J_2}{\partial
u_y}+\left[-\frac{4}{3} T_t u_t-\frac{1}{3} uT_{tt}+\frac{1}{3} x^2
T_{ttt}+\frac{2}{3}xy^2T_{tttt}\right. \nonumber \\ && \left.+2x
yX_{ttt}+2xY_{tt}+\frac{1}{9}y^4T_{ttttt}
+\frac{2}{3}X_{tttt}y^3+2Y_{ttt}y^2
+Z_{1,t}y+Z_{2,t}\right.\nonumber\\&&\left.-(\frac{1}{3} x
T_{tt}+\frac{1}{3}y^2 T_{ttt}+y X_{tt}+Y_{t}) u_x -(\frac{2}{3} y
T_{tt}+X_t)u_y)\right]\frac{\partial J_2}{\partial u_t} +(-T_t
u_{xx}\nonumber \\
&&+\frac{2}{3}T_{tt})\frac{\partial J_2}{\partial
u_{xx}}+\left[-\frac{4}{3} T_t u_{xy}+\frac{4}{3}y T_{ttt}
+2X_{tt}-(\frac{2}{3}yT_{tt}+X_t)u_{xx} \right]\frac{\partial
J_2}{\partial u_{xy}}\nonumber\\
&&+\left[-\frac{5}{3}T_t u_{xt}-\frac{2}{3}T_{tt} u_x+\frac{2}{3}
xT_{ttt}+\frac{2}{3}y^2T_{tttt}+2X_{ttt}y+2Y_{tt}-(\frac{1}{3} x
T_{tt}\right. \nonumber\\
&&\left.+\frac{1}{3} y^2 T_{ttt}+y X_{tt} +Y_t)u_{xx}-(\frac{2}{3}y
T_{tt}+X_t)u_{xy}\right]\frac{\partial J_2}{\partial
u_{xt}}+\left[-\frac{5}{3}T_tu_{yy}\right.\nonumber\\&&\left.+\frac{4}{3}xT_{ttt}+\frac{4}{3}y^2T_{tttt}
+4X_{ttt}y+4Y_{tt}-\frac{2}{3}u_xT_{tt}-2(\frac{2}{3}yT_{tt}+X_t)u_{xy}\right]\frac{\partial
J_2}{\partial
u_{yy}}\nonumber\\
&&+\left[-2u_{yt}T_t-u_yT_{tt}+\frac{4}{3}xyT_{tttt}+2xX_{ttt}+\frac{4}{9}y^3T_{ttttt}
+2X_{tttt}y^2+4Y_{ttt}y+Z_{1,t}\right. \nonumber\\
&&\left.-(\frac{2}{3}yT_{ttt}+X_{tt})u_x-(\frac{2}{3}yT_{tt}
+X_t)u_{tx}-(\frac{1}{3}xT_{tt}+\frac{1}{3}y^2T_{ttt}+X_{tt}y+Y_t)u_{xy}\right. \nonumber\\
&&\left.-(\frac{2}{3}yT_{tt} +X_t)u_{yy}\right]\frac{\partial
J_2}{\partial
u_{yt}}+\left[-\frac{7}{3}u_{tt}T_t-\frac{5}{3}u_tT_{tt}-\frac{1}{3}uT_{ttt}
+\frac{1}{3}x^2T_{tttt}\right.\nonumber\\&&\left.+\frac{2}{3}xy^2T_{ttttt}+2xX_{tttt}y+2xY_{ttt}
+\frac{1}{9}y^4T_{tttttt}+\frac{2}{3}X_{ttttt}y^3+2Y_{tttt}y^2\right. \nonumber\\
&&\left.+Z_{1,tt}y
+Z_{2,tt}-(\frac{1}{3}xT_{ttt}+\frac{1}{3}y^2T_{tttt}+X_{ttt}y+Y_{tt})u_x-2(\frac{1}{3}
xT_{tt}\right. \nonumber\\
&&\left.+\frac{1}{3}y^2T_{ttt}+X_{tt}y+Y_t)u_{tx}-(\frac{2}{3}yT_{ttt}
+X_{tt})u_y-2(\frac{2}{3}yT_{tt}+X_t)u_{yt}\right]\frac{\partial
J_2}{\partial u_{tt}}\nonumber\\&&+\frac{4}{3} T_t
J_2-(\frac{2}{3}yT_{tt}+X_t)\rho=0, \label{cl2}
\end{eqnarray}
\begin{eqnarray}
&&(\frac{1}{3}xT_t+\frac{1}{3}y^2T_{tt}+yX_t+Y)\frac{\partial
\rho}{\partial x} +(\frac{2}{3}y T_t+X )\frac{\partial
\rho}{\partial y} +T\frac{\partial \rho}{\partial t}
+(-\frac{1}{3}uT_t\nonumber \\
&&+\frac{1}{3}x^2T_{tt}+\frac{2}{3}xy^2 T_{ttt}+2xy
X_{tt}+2xY_t+\frac{1}{9}y^4T_{tttt}+\frac{2}{3}X_{ttt}y^3+2Y_{tt}y^2\nonumber \\
&&+Z_1y+Z_2)\frac{\partial \rho}{\partial u}+(-\frac{2}{3}T_t
u_x+\frac{2}{3}xT_{tt}+\frac{2}{3}y^2T_{ttt}+2X_{tt}y+2Y_t)\frac{\partial
\rho}{\partial u_x}\nonumber\\&&+\left[-T_t u_y+\frac{4}{3}x y
T_{ttt}+2x
X_{tt}+\frac{4}{9}y^3T_{tttt}+2X_{ttt}y^2+4Y_{tt}y+Z_1\right.
\nonumber \\ && \left. -(\frac{2}{3} y
T_{tt}+X_t)u_x\right]\frac{\partial \rho}{\partial u_y}
+\left[-\frac{4}{3} T_t u_t-\frac{1}{3} uT_{tt}+\frac{1}{3} x^2
T_{ttt}+\frac{2}{3}xy^2T_{tttt}\right. \nonumber \\ && \left. +2x
yX_{ttt}+2xY_{tt}+\frac{1}{9}y^4T_{ttttt}+\frac{2}{3}X_{tttt}y^3+2Y_{ttt}y^2
+Z_{1,t}y+Z_{2,t}\right.\nonumber\\&&\left.-(\frac{1}{3} x
T_{tt}+\frac{1}{3}y^2 T_{ttt}+y X_{tt}+Y_{t}) u_x -(\frac{2}{3} y
T_{tt}+X_t)u_y)\right]\frac{\partial \rho}{\partial u_t}\nonumber \\
&&+(-T_t u_{xx}+\frac{2}{3}T_{tt})\frac{\partial \rho}{\partial
u_{xx}}+\left[-\frac{4}{3} T_t u_{xy}+\frac{4}{3}y T_{ttt}
+2X_{tt}-(\frac{2}{3}yT_{tt}\right. \nonumber\\
&&\left.+X_t)u_{xx} \right]\frac{\partial \rho}{\partial
u_{xy}}+\left[-\frac{5}{3}T_t u_{xt}-\frac{2}{3}T_{tt}
u_x+\frac{2}{3}
xT_{ttt}+\frac{2}{3}y^2T_{tttt}+2X_{ttt}y\right. \nonumber\\
&&\left.+2Y_{tt}-(\frac{1}{3} x T_{tt}+\frac{1}{3} y^2 T_{ttt}+y
X_{tt} +Y_t)u_{xx}-(\frac{2}{3}y
T_{tt}+X_t)u_{xy}\right]\frac{\partial \rho}{\partial
u_{xt}}\nonumber\\&&+\left[-\frac{5}{3}T_tu_{yy}+\frac{4}{3}xT_{ttt}+\frac{4}{3}y^2T_{tttt}
+4X_{ttt}y+4Y_{tt}-\frac{2}{3}u_xT_{tt}\right.\nonumber\\
&&\left.-2(\frac{2}{3}yT_{tt}+X_t)u_{xy}\right]\frac{\partial
\rho}{\partial
u_{yy}}+\left[-2u_{yt}T_t-u_yT_{tt}+\frac{4}{3}xyT_{tttt}
+2xX_{ttt}\right.\nonumber\\&&\left.+\frac{4}{9}y^3T_{ttttt}
+2X_{tttt}y^2+4Y_{ttt}y+Z_{1,t}-(\frac{2}{3}yT_{ttt}+X_{tt})u_x-(\frac{2}{3}yT_{tt}\right. \nonumber\\
&&\left.
+X_t)u_{tx}-(\frac{1}{3}xT_{tt}+\frac{1}{3}y^2T_{ttt}+X_{tt}y+Y_t)u_{xy}-(\frac{2}{3}yT_{tt}
+X_t)u_{yy}\right]\frac{\partial \rho}{\partial
u_{yt}}\nonumber\\
&&+\left[-\frac{7}{3}u_{tt}T_t-\frac{5}{3}u_tT_{tt}-\frac{1}{3}uT_{ttt}
+\frac{1}{3}x^2T_{tttt}+\frac{2}{3}xy^2T_{ttttt}+2xX_{tttt}y+2xY_{ttt}
\right. \nonumber\\
&&\left.+\frac{1}{9}y^4T_{tttttt}+\frac{2}{3}X_{ttttt}y^3+2Y_{tttt}y^2+Z_{1,tt}y
+Z_{2,tt}-(\frac{1}{3}xT_{ttt}+\frac{1}{3}y^2T_{tttt}\right. \nonumber\\
&&\left.+X_{ttt}y+Y_{tt})u_x-2(\frac{1}{3}
xT_{tt}+\frac{1}{3}y^2T_{ttt}+X_{tt}y+Y_t)u_{tx}-(\frac{2}{3}yT_{ttt}
+X_{tt})u_y\right.\nonumber\\&&\left.-2(\frac{2}{3}yT_{tt}+X_t)u_{yt}\right]\frac{\partial
\rho}{\partial u_{tt}}+T_t \rho=0, \label{cl3}
\end{eqnarray}
The solutions $J_1$, $J_2$ and $\rho$ of (\ref{cl1})-(\ref{cl3}) can
be directly solved:
\begin{eqnarray}
\rho=f_0(t)K_1(t_1, \ t_2, \ t_3, \cdots,\ t_{12}), \label{rho}
\end{eqnarray}
\begin{eqnarray}
J_2&=&[f_1(t)+f_2(t) y]K_1(t_1,\ t_2,\ t_3,\ \cdots,\
t_{12})\nonumber\\&&+f_3(t) K_2(t_1,\ t_2,\ t_3,\ \cdots,\ t_{12}),
\label{j2}
\end{eqnarray}
\begin{eqnarray}
J_1&=& [f_4(t)+f_5(t) x+f_6(t) y+f_7(t) y^2]K_1(t_1,\ t_2,\ t_3,\
\cdots,\ t_{12})\nonumber \\ &&+[f_8(t) +f_9(t) y]K_2(t_1,\ t_2,\
t_3,\ \cdots,\ t_{12})\nonumber\\&&+f_{10}(t) K_3(t_1,\ t_2,\ t_3,\
\cdots,\ t_{12}), \label{j1}
\end{eqnarray}
where $K_1$, $K_2$ and $K_3$ are arbitrary functions of $\{t_1,\
t_2,\ t_3,\ \cdots,\ t_{12}\}$, and $f_i$, $i=0,\ 1,\ \cdots, \ 10$
are functions fixed by:
\begin{eqnarray}
f_0=T^{-1},
\end{eqnarray}
\begin{eqnarray}
f_1=XT^{-2}, \qquad f_2=\frac{2 }{3}T_t T^{-2}, \qquad
f_3=T^{-\frac{4}{3}},
\end{eqnarray}
\begin{eqnarray}
&& f_4=YT^{-2}, \quad f_5=\frac{1}{3}T_{t} T^{-2}, \quad
f_6=X_tT^{-2}, \quad f_7=\frac{1 }{3}T_{tt}T^{-2}, \nonumber
\\ && f_8=XT^{-\frac{7}{3}}, \quad f_9=\frac{2}{3}T_t
T^{-\frac{7}{3}}, \quad f_{10}=T^{-\frac{5}{3}},
\end{eqnarray}
with the invariants being
\begin{eqnarray}
t_1=T^{-\frac{2}{3}}y-X_1,
\end{eqnarray}
\begin{eqnarray}
t_2=T^{-\frac{1}{3}}x-\frac{1}{3} T^{-\frac{4}{3}}T_t y^2-
T^{-\frac{4}{3}} X y-Y_1,
\end{eqnarray}
\begin{eqnarray}
t_3&=&-\frac{1}{3}T^{-\frac{2}{3}}T_tx^2-\frac{10}{81}T^{-\frac{8}{3}}y^4T_t^3
-2T^{-\frac{2}{3}}y^2Y_t-\frac{5}{3}T^{-\frac{8}{3}}yX^3-T^{-\frac{2}{3}}y
Y_2\nonumber \\ &&+T^{-\frac{5}{3}}xX^2-2T^{-\frac{2}{3}}x
Y+\frac{2}{9}T^{-\frac{5}{3}}y^4T_{tt}T_t
+\frac{8}{9}T^{-\frac{5}{3}}y^3X_tT_t\nonumber \\
&&+\frac{4}{9}T^{-\frac{5}{3}}y^3XT_{tt} -2T^{-\frac{2}{3}}y xX_t
+\frac{4}{3}T^{-\frac{5}{3}}y x XT_t-\frac{2}{3}T^{-\frac{2}{3}}x
y^2T_{tt}\nonumber \\
&&-\frac{1}{9}T^{-\frac{2}{3}}y^4T_{ttt}-\frac{2}{3}T^{-\frac{2}{3}}y^3X_{tt}
-\frac{20}{27}T^{-\frac{8}{3}}y^3XT_t^2
+\frac{4}{9}T^{-\frac{5}{3}}x y^2T_t^2\nonumber
\\ &&+2T^{-\frac{5}{3}}y^2XX_t-\frac{5}{3}T^{-\frac{8}{3}}y^2X^2T_t
+\frac{4}{3}T^{-\frac{5}{3}}y^2T_tY+4T^{-\frac{5}{3}}y X Y
\nonumber\\&&+uT^{\frac{1}{3}}+\frac{1}{3}Y_3 ,
\end{eqnarray}
\begin{eqnarray}
t_4&=&u_xT^{\frac{2}{3}}-\frac{2}{3}T^{-\frac{1}{3}}T_tx+\frac{4}{9}T^{-\frac{4}{3}}
y^2T_t^2-\frac{2}{3}T^{-\frac{1}{3}}y^2T_{tt}-2T^{-\frac{1}{3}}yX_t
\nonumber
\\&&+\frac{4}{3}T^{-\frac{4}{3}}yXT_t+X^2T^{-\frac{4}{3}}-2YT^{-\frac{1}{3}},
\end{eqnarray}
\begin{eqnarray}
t_5&=&\frac{2}{3}T_tu_xy+\frac{2}{3}T^{-1}xXT_t+\frac{4}{9}T^{-1}xyT_t^2
-\frac{4}{3}xyT_{tt}+\frac{4}{9}T^{-1}y^3T_tT_{tt}\nonumber
\\&&+\frac{4}{3}T^{-1}y^2X_tT_t
+\frac{2}{3}T^{-1}y^2XT_{tt}-\frac{8}{9}T^{-2}y^2XT_t^2+2T^{-1}yXX_t
\nonumber
\\&&-\frac{4}{3}T^{-2}yX^2T_t+\frac{4}{3}T^{-1}yYT_t+2XT^{-1}Y+Xu_x-2xX_t
-\frac{4}{9}y^3T_{ttt}\nonumber
\\&&-2y^2X_{tt}-4yY_t-\frac{2}{3}X^3T^{-2}+u_yT-Y_2
-\frac{16}{81}T^{-2}y^3T_t^3,
\end{eqnarray}
\begin{eqnarray}
t_6&=&\frac{1}{9}T^{\frac{1}{3}}(9u_tT-9Z_2+9u_yX
+9Yu_x-T_{tttt}y^4-3T_{tt}x^2 -18Y_{tt}y^2\nonumber
\\&&-9Z_1y-18Y_tx+3T_tu
-6X_{ttt}y^3+6u_yT_ty+3T_txu_x-6xy^2T_{ttt}\nonumber
\\&&+3T_{tt}y^2u_x+9yu_xX_t-18xX_{tt}y),
\end{eqnarray}
\begin{eqnarray}
t_7=u_{xx}T-\frac{2}{3}T_t,
\end{eqnarray}
\begin{eqnarray}
t_8&=&\frac{1}{9}T^{-\frac{2}{3}}(9u_{xy}T^2+9u_{xx}TX+6u_{xx}yT_tT
+4yT_t^2-12yTT_{tt}\nonumber\\&&-18TX_t+6XT_t),
\end{eqnarray}
\begin{eqnarray}
t_9&=&\frac{1}{3}T^{\frac{2}{3}}(3u_{tx}T+3u_{xy}X+2u_{xy}yT_t+3u_{xx}Y
+3u_{xx}yX_t-2y^2T_{ttt}\nonumber
\\&&+u_{xx}xT_t+u_{xx}y^2T_{tt}-2xT_{tt}-6Y_t
+2u_xT_t-6X_{tt}y),
\end{eqnarray}
\begin{eqnarray}
t_{10}&=&\frac{1}{27}T^{-\frac{4}{3}}(27u_{yy}T^3+54u_{xy}T^2X+36u_{xy}yT^2T_t
+27u_{xx}X^2T\nonumber \\
&&+36u_{xx}yTXT_t+12u_{xx}y^2T_t^2T-108T^2Y_t-18X^2T_t+36TYT_t
\nonumber \\
&&+18T^2T_tu_x+12xTT_t^2-36xT_{tt}T^2-8y^2T_t^3+12y^2TT_tT_{tt}
\nonumber \\
&&-36y^2T_{ttt}T^2-108yT^2X_{tt}+36yTX_tT_t-24yXT_t^2),
\end{eqnarray}
\begin{eqnarray}
t_{11}&=&u_{ty}T^2-Z_1T+u_{xy}X^2-2XY_t+\frac{4}{9}u_xyT_t^2
+\frac{2}{3}u_xXT_t\nonumber
\\&&-\frac{4}{3}yY_tT_t+u_xTX_t+u_{yy}TX+u_{tx}TX
+\frac{4}{9}u_{xy}y^2T_t^2\nonumber
\\&&-\frac{2}{3}xXT_{tt}-\frac{2}{3}y^2XT_{ttt}
-2yXX_{tt}-\frac{4}{9}y^3T_{ttt}T_t-\frac{4}{3}y^2X_{tt}T_t\nonumber
\\&&-2xTX_{tt}
-2y^2TX_{ttt}-\frac{4}{9}y^3TT_{tttt}-4yTY_{tt}+u_{xy}TY\nonumber
\\&&+u_{xx}X
Y+\frac{1}{3}TT_tu_{xy}x
+u_{xy}yTX_t+\frac{1}{3}u_{xy}y^2TT_{tt}-\frac{4}{3}xyTT_{ttt}\nonumber
\\&&+\frac{2}{3}u_xyTT_{tt}
+u_{xx}yXX_t+\frac{2}{3}u_{xx}yT_tY+\frac{1}{3}u_{xx}y^2XT_{tt}
\nonumber\\&&+\frac{2}{3}u_{xx}y^2T_tX_t+
\frac{4}{3}u_{xy}yXT_t-\frac{4}{9}xyT_{tt}T_t+\frac{2}{9}u_{xx}y^3T_{tt}T_t
\nonumber\\&&+\frac{1}{3}u_{xx}xXT_t+\frac{2}{9}u_{xx}xyT_t^2+\frac{2}{3}TT_tu_{yy}y
+\frac{2}{3}u_{tx}TT_ty+TT_tu_y,
\end{eqnarray}
\begin{eqnarray}
t_{12}&=&\frac{1}{9}T^{\frac{1}{3}}(-9Z_{2,t}T-9Z_1X+uT_t^2+15u_xyX_tT_t
+9u_{tt}T^2\nonumber\\&&+9u_{yy}X^2+9u_{xx}Y^2+15u_tT_tT-18Y_tY+18u_{ty}TX\nonumber\\&&+18u_{tx}T
Y-6y^2T_{tt}Y_t-6y^2T_{ttt}Y+3uTT_{tt}-18y^2TY_{ttt}\nonumber\\&&-6y^3TX_{tttt}-18xTY_{tt}
-Ty^4T_{ttttt}-3Tx^2T_{ttt}-14y^3X_{ttt}T_t\nonumber\\&&-4y^3XT_{tttt}-18xXX_{tt}
-6xT_{tt}Y-18y^2X_{tt}X_t-30y^2Y_{tt}T_t\nonumber\\&&+9u_{xx}y^2X_t^2
-6y^3T_{tt}X_{tt}-6y^3T_{ttt}X_t-2y^4T_{tt}T_{ttt}-3y^4T_tT_{tttt}
\nonumber\\&&-3x^2T_{tt}T_t+u_{xx}y^4T_{tt}+18u_{xy}XY+u_{xx}x^2T_t^2
+4u_{yy}y^2T_t^2\nonumber\\&&-18yYX_{tt}-36yXY_{tt}-2xy^2T_{tt}^2-18y^2XX_{ttt}
-3Z_2T_t\nonumber\\&&+2u_{xx}xy^2T_{tt}T_t+6u_{xx}xyT_tX_t-9yTZ_{1,t}+9u_yTX_t
+9u_xTY_t\nonumber\\&&-12xY_tT_t+12u_yXT_t+9u_xXX_t+9u_xT_tY-18yY_tX_t
-9yZ_1T_t\nonumber\\&&+3u_xxT_t^2+8u_yyT_t^2+9u_xyTX_{tt}+6u_{tx}y^2TT_{tt}
+6u_{tx}xT_tT\nonumber\\&&+12u_{ty}yT_tT-18xyTX_{ttt}+3Tu_xy^2T_{ttt}
+18u_{tx}yTX_t\nonumber\\&&+3u_xxTT_{tt}+6u_yyTT_{tt}-6Txy^2T_{tttt}+12u_{xy}y^2T_tX_t
\nonumber\\&&+6u_{xy}y^2XT_{tt}+6u_xyXT_{tt}+4u_{xy}xyT_t^2+4u_{xy}y^3T_{tt}T_t
\nonumber\\&&+6u_{xy}xXT_t+12u_{yy}yXT_t+6u_{xx}xYT_t+6u_{xx}y^3T_{tt}X_t
\nonumber\\&&+18u_{xx}yYX_t-24xyX_{tt}T_t-6xyX_tT_{tt}-12xyXT_{ttt}\nonumber\\&&+7u_xy^2T_{tt}T_t
-12xy^2T_{ttt}T_t+6u_{xx}y^2YT_{tt}\nonumber\\&&+18u_{xy}yXX_t+12u_{xy}yYT_t),
\end{eqnarray}
where
\begin{eqnarray}
X_{1t}=XT^{-\frac{5}{3}},
\end{eqnarray}
\begin{eqnarray}
Y_{1t}= -T^{-\frac{7}{3}}X^2+T^{-\frac{4}{3}} Y,
\end{eqnarray}
\begin{eqnarray}
Y_{2t}=Z_1,
\end{eqnarray}
\begin{eqnarray}
Y_{3t}=-T^{-\frac{11}{3}}(15YX^2T-6Y^2T^2+3Z_2T^3-5X^4-3XY_2T^2).
\end{eqnarray}

To determine the functions of $K_1$, $K_2$, and $K_3$, we substitute
(\ref{rho}), (\ref{j2}), and (\ref{j1}) into (\ref{cl0}) which
yields a complicated equation
\begin{eqnarray}
&& J_{1,x}+J_{1,u} u_x+J_{1,u_x} u_{xx}+J_{1, u_y} u_{xy}+J_{1,u_t}
u_{xt}+J_{1, u_{xx}}u_{xxx}\nonumber \\ &&+J_{1,
u_{xy}}u_{xxy}+J_{1, u_{xt}} u_{xxt}  +J_{1, u_{yy}} u_{xyy}+J_{1,
u_{yt}} u_{xyt}+J_{1, u_{tt}} u_{xtt}\nonumber \\ &&+J_{2,y}+J_{2,u}
u_y+J_{2,u_x} u_{xy}+J_{2, u_y} u_{yy}+J_{2,u_t} u_{yt}  +J_{2,
u_{xx}}u_{xxy}\nonumber\\&&+J_{2, u_{xy}}u_{xyy}+J_{2, u_{xt}}
u_{xyt}+J_{2, u_{yy}} u_{yyy}+J_{2, u_{yt}} u_{yyt}+J_{2, u_{tt}}
u_{ytt}\nonumber
\\ &&+\rho_{t}+\rho_{u} u_t  +\rho_{u_x} u_{xt}+\rho_{u_y}
u_{yt}+\rho_{u_t} u_{tt}+\rho_{u_{xx}}u_{xxt}
+\rho_{u_{xy}}u_{xyt}\nonumber
\\ &&+\rho_{u_{xt}} u_{xtt}+\rho_{ u_{yy}}
u_{yyt}+\rho_{u_{yt}} u_{ytt}  +\rho_{u_{tt}} u_{ttt}=0.
\label{extendcl}
\end{eqnarray}

To solve the complicated equation (\ref{extendcl}), we begin from
the highest derivatives of $u$ for $K_1$, $K_2$ and $K_3$ being
$\{u_{xxx}, \ u_{xxy}, \ \cdots, \ u_{ttt}\}$ independent. Letting
the coefficients of $\{u_{xxx}, \ u_{xxy}, \ \cdots, \ u_{ttt}\}$ be
zero, we can get a more simplified equation. For example, the term
of $u_{ttt}$ in (\ref{extendcl}) is
\begin{eqnarray}
3 T^{3} K_{1, t_{12}}u_{ttt},
\end{eqnarray}
where $K_{i, t_{j}^n}= \frac{\partial^j K_n}{\partial t_i^j}$. There
is no nontrivial solution for $K_{1, t_{12}}\neq 0$, thus the only
possible case is
\begin{eqnarray}
K_{1, t_{12}}=0, \quad {i. e.}, \quad K_1 \equiv K_1(t_1, \ t_2, \
\cdots, \ t_{11}). \label{K1}
\end{eqnarray}
Under (\ref{K1}), the coefficient of $u_{ytt}$ is
\begin{eqnarray}
3 T^{\frac{8}{3}} (K_{2, t_{12}}+K_{1, t_{11}}),
\end{eqnarray}
which leads to the only possible solution
\begin{eqnarray}
K_2(t_1, \ t_2, \ t_3, \ \cdots, \ t_{12})=-t_{12} K_{1,
t_{11}}+K_{21} (t_1, \ t_2, \ t_3, \ \cdots, \ t_{11}),
\end{eqnarray}
with $K_{21} (t_1,\ t_2,\ t_3,\ \cdots,\ t_{11})$ being an
un-determined function of the indicated variables.

Like the procedure to eliminate $u_{ttt}$ and $u_{ytt}$, vanishing
the terms of $u_{xxx}$, $u_{xxy}$, $\cdots$, $u_{yyy}$ results in
\begin{eqnarray}
K_1&=&(-F_2t_8+F_3t_7+F_{10})t_{11}+(F_1t_7+F_{11}+F_2t_9)t_{10}\nonumber
\\&&+(F_4-t_8F_3)t_9-F_1t_8^2+(F_5+F_6)t_8+F_8t_7+F_{14},
\end{eqnarray}
\begin{eqnarray}
K_2&=&(F_2t_8-F_3t_7-F_{10})t_{12}+(-F_1t_7-F_2t_9-F_{11})t_{11}\nonumber
\\&&+t_9^2F_3
+(F_5+F_1t_8)t_9+F_7t_7+F_9t_8+F_{13},
\end{eqnarray}
\begin{eqnarray}
K_3&=&(-F_2t_{10}+t_8F_3-F_4)t_{12}+F_2t_{11}^2+(F_1t_8-F_3t_9-2F_5\nonumber
\\&&
-F_6)t_{11}+(-F_1t_9-F_9)t_{10}-F_7t_8-F_8t_9+F_{12},
\end{eqnarray}
with $14$ equations satisfied:
\begin{eqnarray}
&&F_{1,t_3}t_6+F_{7,t_5}-F_{9,t_4}=0, \nonumber
\\
&&-F_{2,t_4}+F_{1,t_6}-F_{3,t_5}=0, \nonumber
\\
&&F_{8,t_3}t_6+F_{12,t_4}+F_{7,t_1}+F_{7,t_3}t_5=0,\nonumber
\\
&&-F_{4,t_4}+F_{8,t_6}-F_{3,t_3}t_5-F_{3,t_1}=0, \nonumber \\ &&
-F_{9,t_2}+F_{13,t_5}-F_{9,t_3}t_4+F_{11,t_3}t_6=0, \nonumber
\\&&-F_{11,t_6}+F_{2,t_2}+F_{10,t_5}+F_{2,t_3}t_4=0,
\nonumber
\\ &&
-F_{10,t_1}-F_{4,t_3}t_4+F_{14,t_6}-F_{10,t_3}t_5-F_{4,t_2}=0,\nonumber
\\ &&
F_{13,t_1}+F_{12,t_3}t_4+F_{13,t_3}t_5+F_{14,t_3}t_6+F_{12,t_2}=0,
\nonumber
\\ &&
-F_{5,t_5}-F_{11,t_4}+F_{1,t_3}t_4+F_{1,t_2}-F_{2,t_3}t_6+F_{9,t_6}=0,
\nonumber
\\ &&
F_{14,t_4}+F_{5,t_3}t_5+F_{5,t_1}+F_{12,t_6}-F_{8,t_3}t_4-F_{8,t_2}
+F_{4,t_3}t_6=0, \nonumber
\\ &&
-F_{8,t_5}-F_{7,t_6}+2F_{5,t_4}+F_{1,t_3}t_5-F_{3,t_3}t_6
+F_{6,t_4}+F_{1,t_1}=0,\nonumber \\
&&-F_{4,t_5}+F_{2,t_3}t_5-F_{10,t_4}+F_{2,t_1}+F_{3,t_3}t_4
+F_{6,t_6}+F_{5,t_6}\nonumber\\&&+F_{3,t_2}=0, \nonumber
\\
&&(F_{5,t_3}+F_{6,t_3})t_6+F_{9,t_1}-F_{7,t_3}t_4+F_{12,t_5}
+F_{9,t_3}t_5+F_{13,t_4}\nonumber\\&&-F_{7,t_2}=0, \nonumber
\\
&&F_{10,t_3}t_6-(F_{6,t_3}+2F_{5,t_3})t_4+F_{13,t_6}+F_{14,t_5}
-F_{11,t_1}-2F_{5,t_2}\nonumber\\&&-F_{11,t_3}t_5-F_{6,t_2}=0,
\end{eqnarray}
where $F_i$, $i=1, 2, \cdots 14$ are functions of $\{t_1,\ t_2,\
t_3,\ t_4,\ t_5,\ t_6\}$. Fortunately, the equations are linear
system which is straightforward to solve.

For simplicity, we just list the final solutions:
\begin{eqnarray}\label{con1}
K_{1}&=&(\alpha_{2,t_6}t_7+\alpha_{5,t_6}-\alpha_{1,t_5t_6}t_8)t_{11}+(\alpha_{1,t_5t_6}t_9
+(\alpha_{1,t_4t_5}\nonumber\\&&+\alpha_{2,t_5}+\alpha_{6,t_5t_5})t_7+\alpha_{5,t_5}
+\alpha_{7,t_5t_5}+\alpha_{1,t_2t_5}+t_4\alpha_{1,t_3t_5})t_{10}
\nonumber\\&&+(-\alpha_{2,t_6}t_8+\alpha_{3,t_6})t_9+(-\alpha_{1,t_4t_5}-\alpha_{2,t_5}
-\alpha_{6,t_5t_5})t_8^2\nonumber\\&&+(\alpha_{5,t_4}+\alpha_{8,t_5}-\alpha_{1,t_1t_5}
-t_4\alpha_{2,t_3}-t_5\alpha_{1,t_3t_5}+\alpha_{3,t_5}-\alpha_{2,t_2})t_8
\nonumber\\&&+((\alpha_{6,t_3t_5}+\alpha_{2,t_3}-\alpha_{10,t_4t_4})t_5
+\alpha_{6,t_2t_4}+\alpha_{11,t_4}+t_4\alpha_{6,t_3t_4}\nonumber\\&&+\alpha_{3,t_4}
-\alpha_{7,t_4t_4}+\alpha_{6,t_3}+\alpha_{2,t_1}+\alpha_{6,t_1t_5}+\alpha_{8,t_4})t_7
+(\alpha_{7,t_3t_5}\nonumber\\&&-\alpha_{10,t_2t_4}+\alpha_{15,t_3}-t_4\alpha_{10,t_3t_4}
+\alpha_{10,t_3}+\alpha_{5,t_3})t_5+t_4^2\alpha_{6,t_3t_3}
\nonumber\\&&+(\alpha_{13,t_3}+\alpha_{11,t_3}-\alpha_{7,t_3t_4}+\alpha_{8,t_3}
+\alpha_{3,t_3}+2\alpha_{6,t_2t_3})t_4\nonumber\\&&+\alpha_{17}+\alpha_{5,t_1}+\alpha_{7,t_1t_5}
+\alpha_{3,t_2}+\alpha_{8,t_2}+\alpha_{10,t_1}+\alpha_{15,t_1}
\nonumber\\&&+\alpha_{6,t_2t_2}-\alpha_{7,t_2t_4}+\alpha_{11,t_2}+\alpha_{13,t_2},
\end{eqnarray}
\begin{eqnarray}
K_{2}&=&(\alpha_{1,t_5t_6}t_8-\alpha_{2,t_6}t_7-\alpha_{5,t_6})t_{12}
+(-\alpha_{1,t_5t_6}t_9\nonumber\\&&+(-\alpha_{1,t_4t_5}-\alpha_{2,t_5}-\alpha_{6,t_5t_5}
)t_7-\alpha_{7,t_5t_5}-\alpha_{1,t_2t_5}\nonumber\\&&-t_4\alpha_{1,t_3t_5}-\alpha_{5,t_5}
)t_{11}+t_9^2\alpha_{2,t_6}+(t_8(\alpha_{1,t_4t_5}\nonumber\\&&+\alpha_{2,t_5}
+\alpha_{6,t_5t_5})-t_6\alpha_{1,t_3t_6}+(\alpha_{6,t_3t_5}+\alpha_{2,t_3}
)t_4\nonumber\\&&-\alpha_{7,t_4t_5}+\alpha_{9,t_6}-\alpha_{1,t_3}-\alpha_{5,t_4}
+\alpha_{6,t_2t_5}+\alpha_{4,t_6}\nonumber\\&&+\alpha_{2,t_2})t_9+\alpha_{4,t_5}t_8
+((\alpha_{10,t_4t_4}-\alpha_{2,t_3}-\alpha_{6,t_3t_5}\nonumber\\&&-\alpha_{1,t_3t_4}
)t_6+\alpha_{4,t_4}+\alpha_{9,t_4}+\alpha_{12,t_4})t_7\nonumber\\&&+((-\alpha_{1,t_3t_3}
+\alpha_{10,t_3t_4})t_4-\alpha_{10,t_3}-\alpha_{5,t_3}-\alpha_{7,t_3t_5}
\nonumber\\&&-\alpha_{1,t_2t_3}+\alpha_{10,t_2t_4}-\alpha_{15,t_3})t_6+\alpha_{9,t_2}
+\alpha_{4,t_2}+\alpha_{12,t_2}\nonumber\\&&+(\alpha_{12,t_3}+\alpha_{14,t_3}+\alpha_{9,t_3}
+\alpha_{4,t_3})t_4+\alpha_{14,t_2}\nonumber\\&&-\alpha_{16,t_2}+\alpha_{18},\label{k1231}
\end{eqnarray}
\begin{eqnarray}\label{con3}
K_{3}&=&(-\alpha_{1,t_5t_6}t_{10}+\alpha_{2,t_6}t_8-\alpha_{3,t_6})t_{12}
+t_{11}^2\alpha_{1,t_5t_6}\nonumber\\&&+(t_8(\alpha_{1,t_4t_5}+\alpha_{2,t_5}
+\alpha_{6,t_5t_5})-\alpha_{2,t_6}t_9-t_4\alpha_{6,t_3t_5}\nonumber\\&&+\alpha_{1,t_3}
+\alpha_{7,t_4t_5}+t_6\alpha_{1,t_3t_6}-\alpha_{4,t_6}-\alpha_{6,t_2t_5}
-\alpha_{9,t_6}\nonumber\\&&-\alpha_{3,t_5}+\alpha_{1,t_1t_5}+t_5\alpha_{1,t_3t_5}
-\alpha_{8,t_5})t_{11}+((-\alpha_{1,t_4t_5}\nonumber\\&&-\alpha_{2,t_5}-\alpha_{6,t_5t_5})t_9
-\alpha_{4,t_5})t_{10}+((-\alpha_{2,t_3}-\alpha_{6,t_3t_5}
\nonumber\\&&+\alpha_{10,t_4t_4})t_5-\alpha_{2,t_1}-\alpha_{3,t_4}-\alpha_{6,t_2t_4}
-\alpha_{6,t_3}-\alpha_{8,t_4}\nonumber\\&&+\alpha_{7,t_4t_4}-\alpha_{6,t_1t_5}
-t_4\alpha_{6,t_3t_4}-\alpha_{11,t_4})t_9\nonumber\\&&+((-\alpha_{10,t_4t_4}+\alpha_{2,t_3}
+\alpha_{6,t_3t_5}+\alpha_{1,t_3t_4})t_6-\alpha_{4,t_4}\nonumber\\&&-\alpha_{9,t_4}
-\alpha_{12,t_4})t_8+(t_5\alpha_{1,t_3t_3}-t_4\alpha_{6,t_3t_3}
-\alpha_{6,t_2t_3}\nonumber\\&&-\alpha_{8,t_3}-\alpha_{3,t_3}-\alpha_{10,t_1t_4}
-\alpha_{11,t_3}-\alpha_{13,t_3}+\alpha_{1,t_1t_3}\nonumber\\&&+\alpha_{7,t_3t_4})t_6+(-
\alpha_{9,t_3}-\alpha_{4,t_3}-\alpha_{12,t_3}-\alpha_{14,t_3})t_5\nonumber\\&&-\alpha_{4,t_1}
-\alpha_{12,t_1}-\alpha_{14,t_1}+\alpha_{16,t_1}-\alpha_{9,t_1},
\end{eqnarray}
with $\alpha_i(i=1,2,\cdots, 5)$ being arbitrary functions of
$\{t_1, \ t_2, \ t_3, \ t_4, \ t_5, \ t_6\}$, $\alpha_i(i=6,\
7,\,8)$ being arbitrary functions of $\{t_1, \ t_2, \ t_3, \ t_4, \
t_5\}$, $\alpha_9$ being arbitrary function of $\{t_1, \ t_2, \ t_3,
\ t_4, \ t_6\}$, $\alpha_i(i=10, \ 11, \ 12)$ being arbitrary
functions of $\{t_1, \ t_2, \ t_3, \ t_4\}$, $\alpha_{13}$,
$\alpha_{14}$, $\alpha_{15}$ being arbitrary functions of $\{t_1, \
t_2, \ t_3\}$, $\alpha_{16}$, $\alpha_{17}$ being arbitrary
functions of $\{t_1, \ t_2\}$ and $\alpha_{18}$ is function of
$t_2$.

Substituting (\ref{con1})-(\ref{con3}) into (\ref{rho})-(\ref{j1}),
we can obtain the conservation laws of the Lin-Tsien equation
associated with the Lie-B\"{a}cklund generator $X_0$. We have
verified that the conserved vector $(J_1, \ J_2, \ \rho)$ really
satisfied Eq. (\ref{cl0}).

\section{Conclusion and discussion}
In this paper, by applying the modified CK's direct method, we set
up Theorem 1, which shows the relationship between new exact
solutions and old ones of the (2+1)-dimensional Lin-Tsien equation.
Using the transformation relations we then get the corresponding KMV
symmetry algebra and the Lie point symmetry which coincide with the
result generated from the standard Lie approach.

We generate the conservation laws of the Lin-Tsien equation related
to the infinite dimensional KMV symmetry group by use of the
Lie-B\"{a}cklund generator up to the second order group invariants.
The existence of arbitrary functions of the group invariants proves
the Lin-Tsien equation has infinitely many conservation laws which
connect with the general Lie point symmetry (\ref{V}). Though the
symmetries and conservation laws are obtained from the Lin-Tsien
equation, the conservation laws we derived is only depended on the
symmetry which may be possessed by many equations.

\end{document}